\documentclass[aps,prl,twocolumn,superscriptaddress,groupedaddress,10pt]{revtex4}

\usepackage{amsmath}
\usepackage{amssymb}
\usepackage{graphicx}
\usepackage{epsfig}
\usepackage{dcolumn}
\usepackage{bm}
\usepackage{bm}
\usepackage{blindtext}
\usepackage{natbib}
\usepackage{booktabs}
\usepackage{mathrsfs}
\usepackage{epstopdf}
\usepackage{color}
\usepackage[colorlinks=true,linkcolor=blue,citecolor=blue]{hyperref}
\usepackage{hyperref}
\usepackage[normalem]{ulem}

\def\be{\begin{equation}}
\def\ee{\end{equation}}
\def\bea{\begin{eqnarray}}
\def\eea{\end{eqnarray}}

\begin{document}

\title{NMR Evidence of Antiferromagnetic Spin Fluctuations in Nd$_{0.85}$Sr$_{0.15}$NiO$_2$}

\author{Yi Cui}
\thanks{These authors contributed equally to this study.}
\affiliation{Department of Physics and Beijing Key Laboratory of Opto-electronic Functional Materials $\&$ Micro-nano Devices, Renmin University of China, Beijing, 100872, China}

\author{C. Li}
\thanks{These authors contributed equally to this study.}
\affiliation{Department of Physics and Beijing Key Laboratory of Opto-electronic Functional Materials $\&$ Micro-nano Devices, Renmin University of China, Beijing, 100872, China}

\author{Q. Li}
\thanks{These authors contributed equally to this study.}
\affiliation{National Laboratory of Solid State Microstructures and Department of Physics, Center for Superconducting Physics and Materials, Collaborative Innovation Center for Advanced Microstructures, Nanjing University, Nanjing 210093, China}

\author{Xiyu Zhu}
\affiliation{National Laboratory of Solid State Microstructures and Department of Physics, Center for Superconducting Physics and Materials, Collaborative Innovation Center for Advanced Microstructures, Nanjing University, Nanjing 210093, China}

\author{Z. Hu}
\affiliation{Department of Physics and Beijing Key Laboratory of Opto-electronic Functional Materials $\&$ Micro-nano Devices, Renmin University of China, Beijing, 100872, China}

\author{Yi-feng Yang}
\affiliation{Beijing National Laboratory for Condensed Matter Physics and Institute of Physics, Chinese Academy of Sciences, Beijing 100190, China}

\author{J.~S.~Zhang}
\affiliation{Mathematics and Physics Department, North China Electric Power University, Beijing 102206, China}

\author{Rong Yu}
\email{rong.yu@ruc.edu.cn}
\affiliation{Department of Physics and Beijing Key Laboratory of Opto-electronic Functional Materials $\&$ Micro-nano Devices, Renmin University of China, Beijing, 100872, China}

\author{Hai-Hu Wen}
\email{hhwen@nju.edu.cn}
\affiliation{National Laboratory of Solid State Microstructures and Department of Physics, Center for Superconducting Physics and Materials, Collaborative Innovation
Center for Advanced Microstructures, Nanjing University, Nanjing 210093, China}

\author{Weiqiang Yu}
\email{wqyu\_phy@ruc.edu.cn}
\affiliation{Department of Physics and Beijing Key Laboratory of Opto-electronic Functional Materials $\&$ Micro-nano Devices, Renmin University of China, Beijing, 100872, China}



\begin{abstract}

Despite the recent discovery of superconductivity in Nd$_{1-x}$Sr$_{x}$NiO$_2$ thin films, the absence of superconductivity and antiferromagnetism in their bulk materials remain a puzzle. Here we report the
$^{1}$H NMR measurements on powdered Nd$_{0.85}$Sr$_{0.15}$NiO$_2$ samples by taking advantage of the enriched proton concentration after hydrogen annealing.
We find a large full width at half maximum of the spectrum, which keeps increasing with decreasing the temperature and exhibits an upturn behavior at low temperatures. The spin-lattice relaxation rate $1/^{1}T_1$ is strongly enhanced when lowering the temperature, developing
a broad peak at about 40~K, then decreases following a spin-wave-like behavior $1/^{1}T_1{\sim}T^2$ at lower temperatures.
These results evidence a short-range glassy antiferromagnetic ordering of magnetic moments below 40~K and dominant antiferromagnetic fluctuations extending to much higher temperatures. Our findings reveal the strong electron correlations in bulk Nd$_{0.85}$Sr$_{0.15}$NiO$_2$, and shed light on the mechanism of superconductivity observed in films of nickelates.

\end{abstract}

\maketitle


The recent discovery of superconductivity in films of the infinite layer nickelate Nd$_{1-x}$Sr$_{x}$NiO$_2$~\cite{Li_Nature_2019,Li_PRL_2020}
with a maximal transition temperature $T_c$ of 14.9~K has attracted a lot of research attention. Nd$_{1-x}$Sr$_{x}$NiO$_2$ has the same lattice structure as the infinite-layer cuprate superconductor (Sr$_{1-x}$Ca$_x$)$_{1-y}$CuO$_2$~\cite{Azuma_Nature_1992}. In the undoped NdNiO$_2$, the Ni$^{1+}$ ion has an electron configuration of
3$d^{9}$ with a single hole in the $d_{x^2-y^2}$ orbital, as Cu$^{2+}$ in cuprate superconductors.
It was proposed that upon Sr$^{2+}$ substitution for Nd$^{3+}$ the nickelate behaves
as a doped Mott insulator, in analogy to the hole-doped cuprates~\cite{Anisimov_PRB_1999}.
However, a number of major differences between these two systems have been noted.
The charge transfer gap between the Ni 3$d$ to O 2$p$ orbitals was estimated to be much larger than that in cuprates~\cite{Fujimori_JCCM_2019,Botana_PRX_2020}.
So it is under debate whether the hole introduced by Sr doping would reside on the O site as in cuprates~\cite{Ku_arXiv_2020}, or on the Ni site forming a 3d$^8$ configuration. The latter could cause an $S=1$ high-spin configuration of Ni that involves multiorbital physics~\cite{Botana_PRX_2020, Ryee_PRB_2020,Wu_PRB_2020,Karp_arxiv_2020,Zhang_PRB_2020,Sawatzky_arXiv_2020,Norman_Physics_2020}. Moreover, the larger charge transfer gap leads to a smaller exchange energy~\cite{Bocquet_PRB_1996,Jiang_PRL_2020}, which may substantially suppress the antiferromagnetism in the nickelate.
The $5d$~\cite{WSLee_NM_2020,Lee_PRB_2004} and $4f$~\cite{Choi_PRB_2020} electrons in Nd$^{3+}$, and/or
interstitial $s$ electrons~\cite{Gu_Commun Phys_2020} may also participate to the low-energy electronic structure.

The above concerns make an open question in relation to the doped Mott insulator scenario of the nickelate and stimulates alternative theoretical proposals involving Kondo effect~\cite{Zhang_PRB_2020,Sawatzky_arXiv_2020}, anti-Kondo effect~\cite{Choi_PRB_2020}, multi-orbital effect~\cite{Chaloupka_PRL_2008,Savrasov_PRB_2018,Mandal_arxiv_2020}, Hund's coupling~\cite{Kotliar_PRB_2020}, and disorder effect.
Surprisingly, more recent measurements on the bulk materials show an absence of superconductivity~\cite{Li_Commun Mater_2020} and no long-range order antiferromagnetism~\cite{Hayward_JPSJ_1999,Hayward_Solid State Sciences_2003,Wang_PRM_2020}. These results not only challenge the similarity between nickelates and cuprates, but also raise an intriguing question: what could be the critical ingredients manipulating the superconductivity in film and bulk nickelates.

Both the Nd$_{1-x}$Sr$_{x}$NiO$_2$ films and the powered bulk crystals are made from
Nd$_{1-x}$Sr$_{x}$NiO$_3$ by post annealing in hydrogen atmosphere to
remove the apical oxygen in the lattice~\cite{Hayward_JPSJ_1999,Hayward_Solid State Sciences_2003,Li_PRL_2020,Li_Commun Mater_2020}.
However, the resistivity of the powdered bulk crystals show an insulating behavior
from room temperature down to 1.5~K.
One may wonder the role of hydrogen annealing: besides removing the apical oxygens,
could it be a key factor of tuning the superconductivity by introducing effective charge doping or inhomogeneity?

Experimental studies on the electronic structure and magnetism are much demanded to address the above questions and to get a convergent understanding of the superconductivity in general~\cite{Shao_CPL_2019,Wu_CPL_2020,Jia_CPL_2020,Liu_CPL_2020}. NMR is an ideal probe of low-energy physics in condensed matter systems, and the hydrogen annealing
motivates us to search for $^{1}$H NMR in bulk Nd$_{0.85}$Sr$_{0.15}$NiO$_2$ samples, whose film counterpart is near optimal doping~\cite{Li_PRL_2020}.

In this paper, we show that the $^{1}$H NMR intensity of Nd$_{0.85}$Sr$_{0.15}$NiO$_2$
is much larger than that of Nd$_{0.85}$Sr$_{0.15}$NiO$_3$, which implies a higher concentration
of hydrogen after annealing. More strikingly, although the hyperfine shift $^{1}K_s$ in the annealed sample is very small,
a large increase of full width at half maximum (FWHM) of the spectrum
is observed in Nd$_{0.85}$Sr$_{0.15}$NiO$_2$.
The spin-lattice relaxation rate is very high, and is largely enhanced with decreasing the temperature, developing a broad peak at about 40~K.
These results signal dominant antiferromagnetic (AFM) fluctuations
in the system. Below 40 K, the FWHM of the spectrum increases dramatically
with an upturn behavior, and the $1/^{1}T_1{\sim}T^2$ showing a spin-wave-like behavior. These suggest a short-range or glassy
AFM order developing below 40~K. The observed enhanced AFM fluctuations and the short-range AFM ordering
are inherent in the strong electron correlations in the bulk materials, and are crucial for the understanding of the magnetism and superconductivity in nickelates.



\begin{figure}[t]
\includegraphics[width=8.5cm]{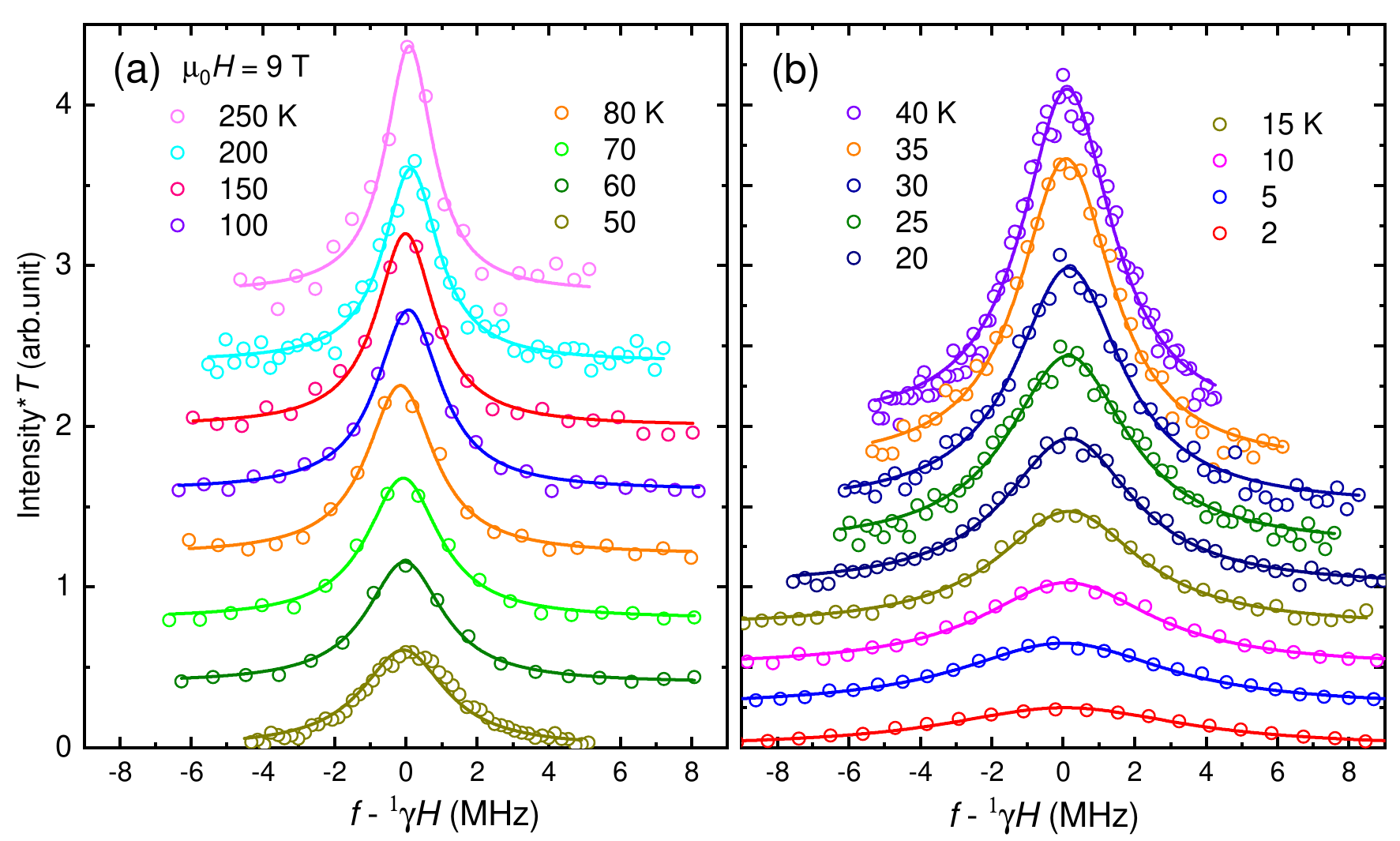}
\caption{\label{spec} The $^{1}$H NMR spectra (open circles) of Nd$_{0.85}$Sr$_{0.15}$NiO$_2$ at a fixed field of $\mu_{0}H$=~9~T,
obtained by frequency sweep at selected temperatures.
The NMR intensity is obtained from the integration of the spin-echo spectrum
multiplying by the temperature. For clarity, vertical shifts are applied to data at different temperatures,
and vertical scales in panels (a) and (b) are chosen differently. The solid lines are function fits to the
spectra with the single-Lorentzian form.}
\end{figure}

The powdered samples of Nd$_{0.85}$Sr$_{0.15}$NiO$_2$ in this study are made
by the topotactic hydrogen annealing method as reported in earlier papers~\cite{Li_Commun Mater_2020,Li_arxiv_2006_10988}.
The magnetization data are measured by a VSM in a Quantum Design PPMS (physical property measurement system).
In principle it is hard to perform NMR measurements on Nd$_{0.85}$Sr$_{0.15}$NiO$_2$ without
isotope enrichment. But after annealing, hydrogen can be injected to the system given its small size. Taking this advantage,
$^{1}$H (spin-1/2, $\gamma$~=~42.5759~MHz/T) NMR signal is found by using a standard spin-echo pulse sequence
$\pi/2$-$\tau$-$\pi$, where $\pi/2$ and $\pi$ pulse lengthes
are 1.5~$\mu$s and 3.0~$\mu$s, respectively, and $\tau$ $\sim$ 1~$\mu$s is the blank time between two
pulses. Such short pulses and $\tau$ are selected to minimize the $^{1}T_2$ effect which
is found to be about 20~$\mu$s at low temperatures. The full NMR spectrum is obtained
by frequency sweep at a fixed field. The spin-lattice relaxation rate
$^{1}T_{1}$ is measured by the inversion-recovery method, and the spin-recovery is fitted
by stretched exponential functions, with a stretching factor $\beta\sim$0.7 at temperature above 50~K and is reduced to about 0.3 when cooled down to 5~K, indicating
either magnetic anisotropy or sample inhomogeneity. In addition to this short $^{1}T_1$ NMR signal,
a long $^{1}T_1$ component with a narrow bandwidth (about 100~kHz) is also found, and does not vary with sample, which should be from the
environment. Considering the single-Lorenzian lineshape, we believe that the spectra in Fig.~\ref{spec} represent the majority phase of the sample, instead of an impurity phase.


\begin{figure}[t]
\includegraphics[width=8.5cm]{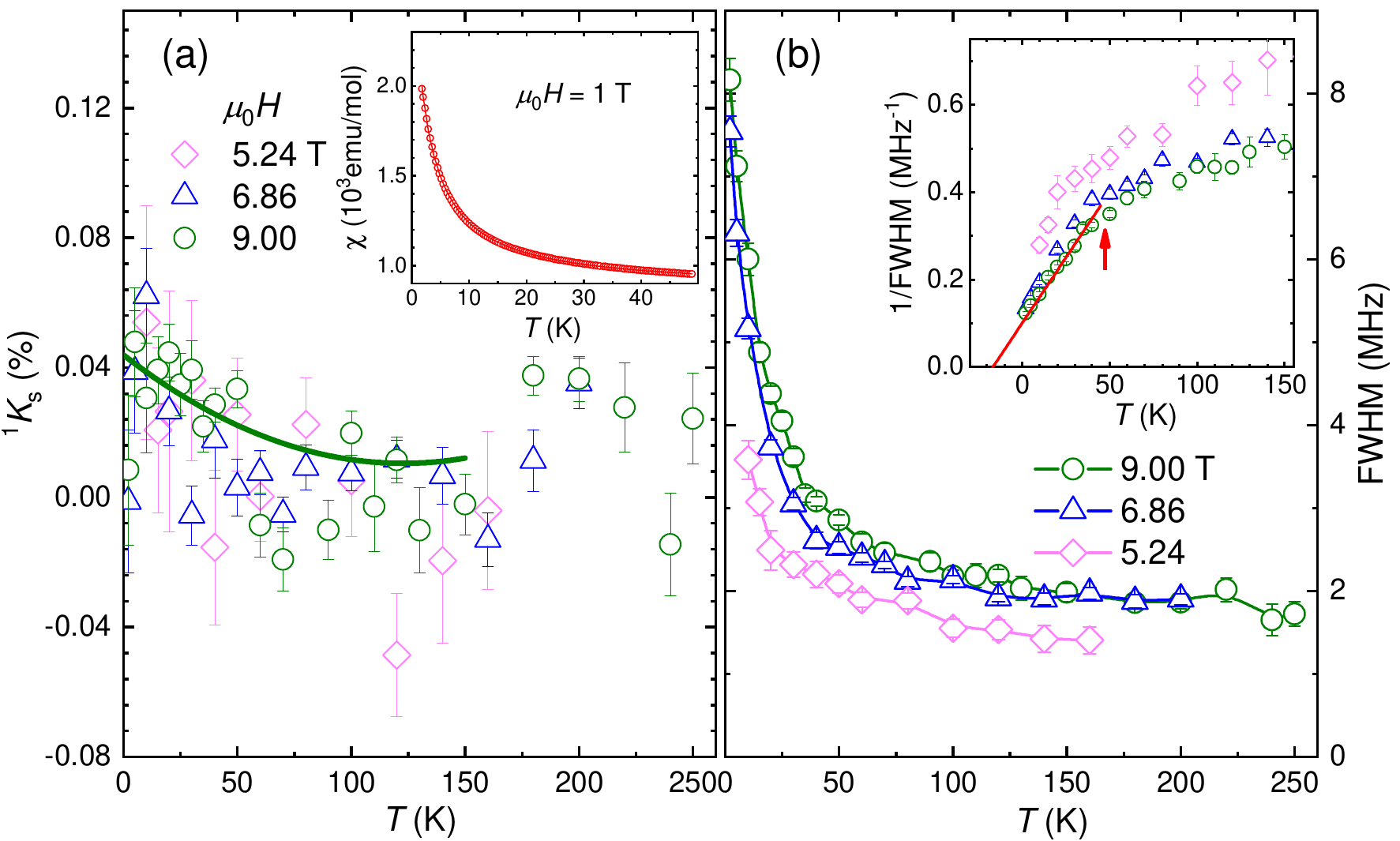}
\caption{\label{kn} (a) The NMR hyperfine shift of Nd$_{0.85}$Sr$_{0.15}$NiO$_2$
as functions of temperatures at different fields. The solid line is a polynomial fit
to the 9~T data below 150 K. Inset: The susceptibility as a function
of temperature with a fixed field $\mu_0H$~=~1~T.
(b) The FWHM of the spectra of Nd$_{0.85}$Sr$_{0.15}$NiO$_2$
as functions of temperatures. Inset: The inverse of the FWHM, which exhibits a
downturn at 40~K marked by the up arrow and a linear temperature dependence below 40~K as shown by the straight line.
}
\end{figure}

The $^{1}$H NMR spectra measured by frequency sweep under a constant field 9~T and various temperatures are
shown in Fig.~\ref{spec}, where zero frequency corresponds to
zero Knight shift. Upon cooling, the NMR spectra follow the Lorentzian form and
barely shift with temperature.
The hyperfine shift $K_s$ (defined as $K_s$~=~$f_{\rm p}/{\gamma}H-1$, where
$f_{\rm p}$ is the peak frequency of the spectrum) and the FWHM are then
calculated from the spectra. Their temperature evolutions at several fields are shown  in Fig.~\ref{kn}(a) and Fig.~\ref{kn}(b), respectively.
$K_s$ is very small, with a typical value less than 0.1\%. Yet a weak upturn
upon cooling is still resolved, which is consistent with the increase of the measured bulk susceptibility
data (inset of Fig.~\ref{kn}(a)).
The small $K_s$ 
values 
and the upturn 
behavior 
barely depend on the field. On the other hand, as shown in Fig.~\ref{kn}(b) the FWHM
is much larger than the hyperfine shift (relative to
the zero frequency), as shown in Fig.~\ref{spec}, and increases with decreasing temperature from 250~K to 40~K, indicating a large magnetic
broadening of the spectra.

Below 40~K, the FWHM increases rapidly with decreasing the temperature and follows an upturn behavior, as also shown by a
kinked downturn (marked by the up arrow) in the inverse of the FWHM (Fig.~\ref{kn}(b) inset).
Such a large broadening is a typical character of inhomogeneous
NMR linewidth, as is much bigger than the measured $1/T_2$ ($T_2\approx$~20~$\mu$s at low  temperatures).
This large linewidth and the upturn behavior are observed over a broad field range.
Though the powdered sample is chemically and structurally inhomogeneous, the strong temperature dependence of the FWHM below 40~K
at large fields can only be attributed to a precursor to static AFM ordering,
because in an antiferromagnet, the projection of local hyperfine fields along the external field
is also randomly distributed between negative and positive values
when the crystalline orientations are random in powders.
This is further discussed below.


\begin{figure}[t]
\includegraphics[width=8.5cm]{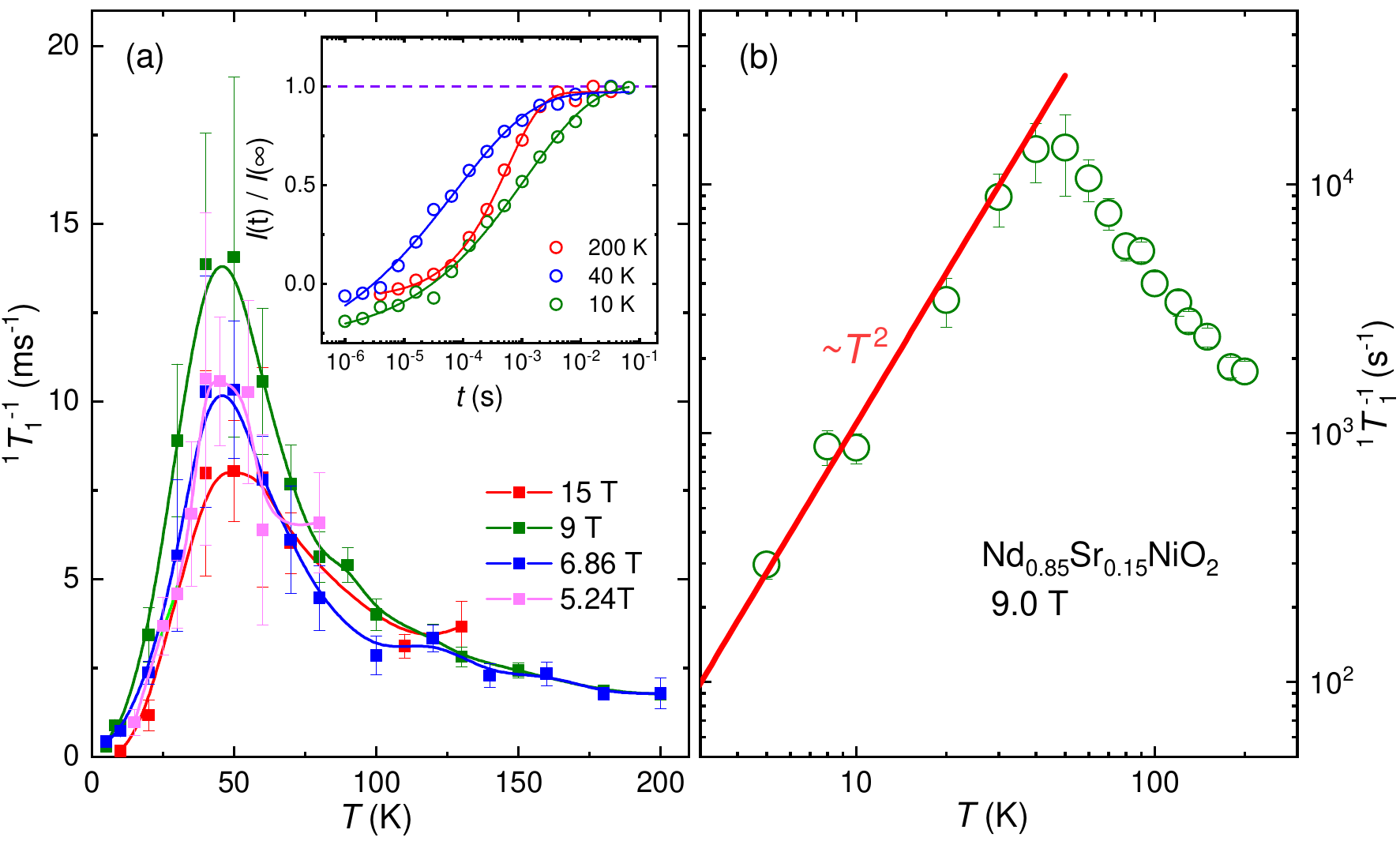}
\caption{\label{invt1t} (a) Measured $1/^1T_1$ with temperature of Nd$_{0.85}$Sr$_{0.15}$NiO$_2$ under different fields.
Inset: The spin recovery data at three typical temperatures, and
the solid lines are fits by the stretched exponential functions to obtain the $T_1$.
(b) The low-temperature $1/^{1}T_1$ plotted in a log-log scale. The solid line is a function fit to
$1/T_1\sim T^2$.
}
\end{figure}

The spin-lattice relaxation rate $1/T_1$, defined as, $1/T_1  = T{\rm lim} _{\omega \to 0} \sum_{q} A_{\rm hf}(q) \, {\rm Im} \chi(q,\omega)/\omega$, where
$\chi(q, \omega)$ is the dynamical susceptibility and $A_{\rm hf}(q)$ the hyperfine coupling,
is a sensitive probe of low-energy spin fluctuations~\cite{Moriya_JPSJ_1963}. Since $^1$H nuclear spin is 1/2,
the spin-lattice relaxation rate $1/^{1}T_1$ of $^{1}$H is only sensitive to magnetic fluctuations.
The $1/^{1}T_1$ is then measured and plotted
as functions of temperatures under various fields in Fig.~\ref{invt1t}.
Upon warming from 2~K, the $1/^{1}T_1$ first increases dramatically, then forms a peaked feature at $T\approx$~40~K,
and finally flattens out at temperatures above 100~K.
This peaked behavior is demonstrated by the spin-recovery curve for the $T_1$, as shown in the inset of Fig.~\ref{invt1t}(a),where the spin recovers
more rapidly at 40~K than that at 200~K and 10~K.

A sharp peak in $1/^{1}T_1$ is usually a signature of magnetic phase transition.
Although the magnetization data revealed a spurious ferromagnetic phase in both
Nd$_{0.85}$Sr$_{0.15}$NiO$_2$~\cite{Li_Commun Mater_2020} and Nd$_{4}$Ni$_3$O$_8$~\cite{Li_PRB_2020, Li_arxiv_2006_10988},
its high onset temperature (above 300~K) are incompatible with our observation.
Since no additional magnetic hysteresis emergent at 40~K by any other probes~\cite{Li_Commun Mater_2020}, the magnetic transition has to be an AFM type.

The broad peak at about 40~K in bulk Nd$_{0.85}$Sr$_{0.15}$NiO$_2$ suggests
the onset of glassy antimagnetic order, where the transition temperature
varies across the sample due to inhomogeneity. This glassy behavior is similar to underdoped cuprate superconductor~\cite{Kiefl_PRL_1989}.
Nevertheless, it indicates the onset of low-energy spin fluctuations at 40~K. Below 40~K, $1/^{1}T_1{\sim}T^\alpha$,
follows a power-law behavior with a low power-law exponent $\alpha$~=~2, much smaller than that caused by AFM spin
wave excitations ($\alpha$ =~5)~\cite{Abragam}, again evidences remaining spin excitations. These results, together with
the broad peak in $1/^{1}T_1$ at 40~K and the large FWHM below 40~K, are fully consistent with a glassy AFM order
caused by inhomogeneity at low temperatures.


\begin{figure}[t]
\includegraphics[width=8.5cm]{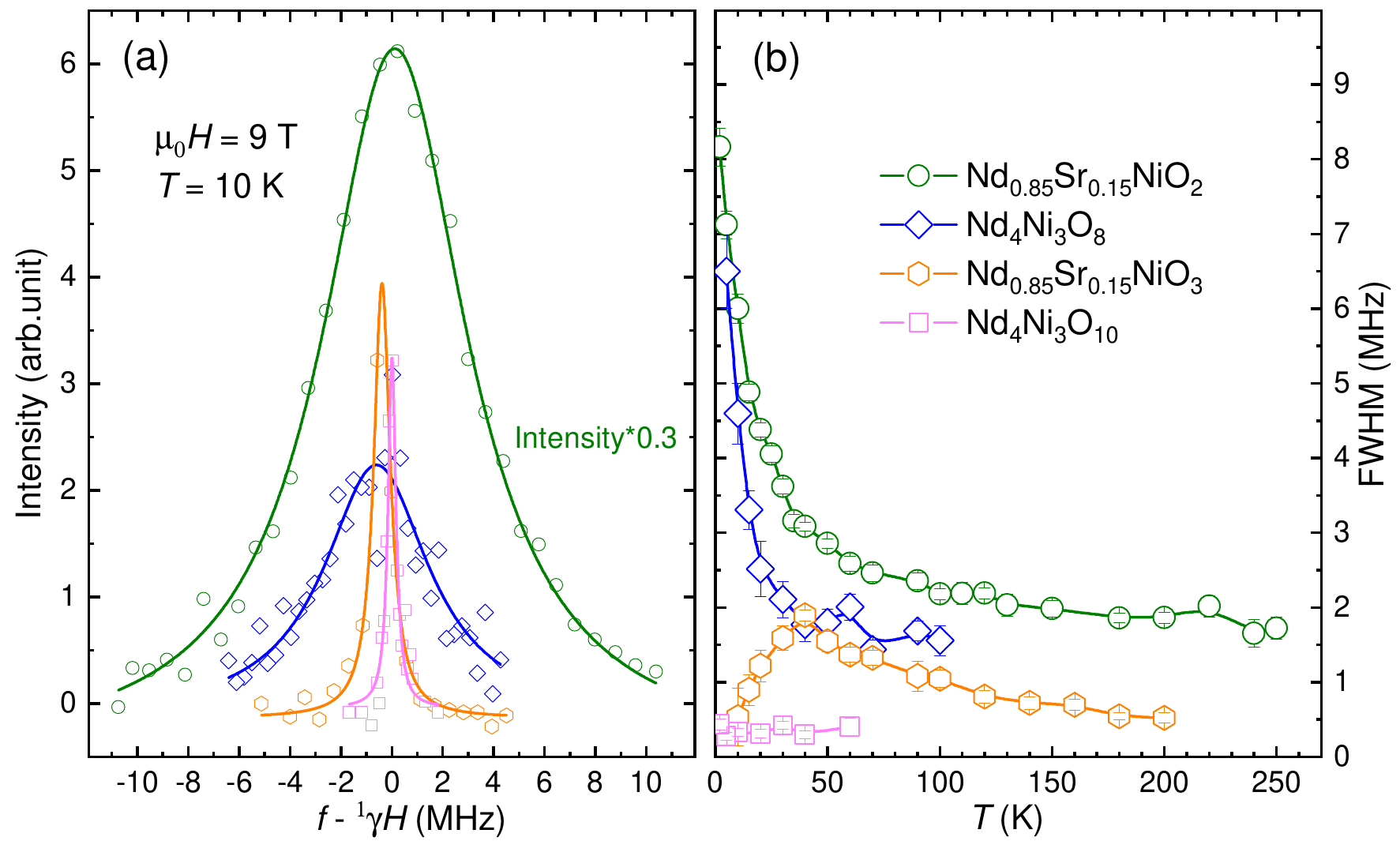}
\caption{\label{fwhm} (a) The $^1$H spectra under a field of 9~T measured on four
compounds of nickelates. The spectra are measured under identical condition, such as  the same sample mass,
coil, NMR pulse sequences, {\it etc.}. Solid lines are Lorentzian-form fit to the data.
(b) Corresponding FWHM as functions of temperatures, obtained by the Lorentzian-form fit to the spectra.
}
\end{figure}

In order to understand the effects of hydrogen annealing and the origin of the $^{1}$H signal, we performed comparative studies on
four bulk nickelate compounds. Nd$_{0.85}$Sr$_{0.15}$NiO$_3$~\cite{Alonso_JSSC_1995} and Nd$_{4}$Ni$_3$O$_{10}$~\cite{Li_arxiv_2006_10988} are made
without hydrogen annealing, and Nd$_{0.85}$Sr$_{0.15}$NiO$_2$ and Nd$_{4}$Ni$_3$O$_8$ are synthesized after hydrogen annealing
from the former two compounds~\cite{Retoux_JSSC_1998,Li_arxiv_2006_10988}, respectively.
Although NdNiO$_3$ is insulating and antiferromagnetically ordered below 200~K~\cite{Scagnoli_PRB_2008,Caviglia_PRB_2013,Kumar_PRB_2013,Hooda_PhysB_2016},
Nd$_{0.85}$Sr$_{0.15}$NiO$_3$ shows a metallic behavior at low temperatures~\cite{Alonso_JSSC_1995}. Nd$_{4}$Ni$_3$O$_{10}$ is {also} metallic
at low temperatures~\cite{Olafsen_JSSC_2000,Li_PRB_2020, Li_arxiv_2006_10988}, but Nd$_{4}$Ni$_3$O$_{8}$~\cite{Li_arxiv_2006_10988}
shows an insulating behavior at low temperatures and its magnetic properties remain unknown.

The spectra of the four compounds at a representative field 9 T are shown in Fig.~\ref{fwhm}(a). Interestingly,
clear $^1$H signals are resolved for the two samples without hydrogen annealing. This is not too
surprising because hydrogen is hard to avoid in materials, given its small atomic size. However,
the total integrated spectral weight of $^{1}$H in Nd$_{0.85}$Sr$_{0.15}$NiO$_2$ is enhanced by a factor
of $\sim$45 compared with Nd$_{0.85}$Sr$_{0.15}$NiO$_3$,
which suggests much higher hydrogen concentration in the annealed sample.
However, by comparing to some organic materials, the concentration of hydrogen in Nd$_{0.85}$Sr$_{0.15}$NiO$_2$ is estimated to be less than 1\%.
Though largely enhanced after annealing, the level of hydrogen concentration remains small and can not provide an effective carrier doping.
We also tried proton injection by using the ionic liquid method~\cite{Cui_SciBull_2018, Cui_CPL_2019,Wei_SciBull_2020},
but no additional effect was observed, implying a similar hydrogen concentration.
On the other hand, the measured NMR spin-lattice relaxation rate takes a typical value
about 1000~s$^{-1}$ for temperatures between 10~K and 200~K, which is much larger than that of the environmental protons,
indicating that the hydrogen ions strongly couple to the magnetic moment in the sample.
Therefore, instead of providing effective doping, hydrogen entered in the system serves as a sensitive local
probe for the magnetic properties of the system.

The respective FWHM of the $^1$H NMR spectra of four compounds are shown in Fig.~\ref{fwhm}(b). For the annealed compounds,
the FWHM are very large. For example, in Nd$_{0.85}$Sr$_{0.15}$NiO$_2$ the FWHM above 50~K corresponds to 0.5\% of the resonance frequency, which is over twenty
times of the hyperfine shift ($\sim$0.02\%). The FWHM of Nd$_{0.85}$Sr$_{0.15}$NiO$_2$ and Nd$_{4}$Ni$_3$O$_8$
are much larger than those of unannealed samples.
In particular, the FWHM in Nd$_{4}$Ni$_3$O$_8$ shows a large upturn below 40~K,
similar to Nd$_{0.85}$Sr$_{0.15}$NiO$_2$. The larger FWHM reflects stronger inhomogeneity of the annealed
compounds. The origin of the inhomogeneity is two-fold. On one hand,
during annealing, the removal of apical oxygen may be incomplete, and
the hydrogen and Ni ions may even form Ni-H bond~\cite{Si_PRL_2020}. These can
certainly increase the level of inhomogeneity in the system. Inhomogeneity promotes
the localization of electrons, which explains the observed insulating behavior up
to room temperature in the annealed samples. On the other hand, the prominent
upturn of FWHM below 40~K in the annealed samples should be associated with
magnetic fluctuations, because at low temperatures the charge and structural
inhomogeneities are quenched and can not account for such a strong temperature dependence.

In the following, we discuss the nature of the observed AFM fluctuations in bulk Nd$_{0.85}$Sr$_{0.15}$NiO$_2$.
At present, we cannot differentiate whether the magnetic fluctuations are caused by Nd or Ni moments.
Nonetheless, the Nd magnetic ordering can be induced by the Ni magnetic ordering through their couplings~\cite{Scagnoli_PRB_2008,Liu_npjQM_2020}.
In either case, our result indicates the existence of intrinsic order of Ni moments.
Consider the dipolar hyperfine coupling between one Ni/Nd moment and the neighbouring $^1$H nuclei (assuming in the apical positions), the FWHM of 4~MHz at 5.24~T gives an estimation of magnetic moment about 0.2~$\mu _B$, which is near the limit of resolution of neutron diffraction and may not be observed~\cite{Hayward_JPSJ_1999,Hayward_Solid State Sciences_2003,Wang_PRM_2020}.
The broad peak of $1/^1T_1$ at around 40~K indicates the
strength of magnetic couplings is at the order of 10~meV.
Indeed, the super-exchange
coupling among nearest Ni moments is estimated theoretically to be 10-30 meV~\cite{Fujimori_JCCM_2019,Botana_PRX_2020,Bocquet_PRB_1996,Jiang_PRL_2020},
consistent with our observations.

Our results support the existence of antiferromagnetically interacting Ni moments and are consistent with the doped Mott insulator scenario of nickelates.
The onset of short-range ordering at 40~K coincides with the upturn of the low-temperature resistivity~\cite{Li_Commun Mater_2020},
which suggests an enhanced insulating behavior through magnetic scattering, as seen in quasi-1D Bechgaard salts~\cite{Lebed}.
A recent RIXS measurement suggests the doped hole results in a low-spin state of the Ni $3d^8$ configuration~\cite{Rossi_arXiv_2020}, which is also consistent
with this picture. Nonetheless, the
alternative picture, in which the doped hole leads to a high-spin $3d^8$ Ni configuration~\cite{Zhang_Vishwanath_PRR_2020,Kotliar_PRB_2020},
cannot be differentiated from our current data.
Measurements on how the AFM fluctuations and short-range order evolve
with different rare earth and/or hole doping would tell more information on whether a low-spin or a high-spin configuration is favored.
As an interesting proposal, the Ni$^{+1}$ moments
could hybridize with Nd $5d$ itinerant electrons to form Kondo singlets~\cite{Zhang_PRB_2020}.
In Nd$_{0.85}$Sr$_{0.15}$NiO$_2$, we do not observe
an effect of Kondo-singlet formation which would cause a
constant $1/T_1T$ at low temperatures~\cite{Hewson}. A likely picture could be the Nd $5d$
bands are already emptied by the hole doping of Sr$^{2+}$ at $x=0.15$, and the Kondo effect
would be more relevant in the underdoped regime. Interestingly, the FWHM with temperature
of Nd$_{0.85}$Sr$_{0.15}$NiO$_3$ shows a peak at about 40~K and drops rapidly at
low temperatures (Fig.~\ref{fwhm}(b)). Whether the Kondo physics is relevant there
deserves further study.

Finally, we discuss the implication to superconductivity of the nickelate.
The observation of AFM fluctuations and short-range magnetic ordering
in the bulk materials suggests that the superconductivity in the nickelate is mediated via spin fluctuations,
analogous to cuprate and iron-based superconductors. In the bulk materials, the absence of
superconductivity and long-range antiferromagnetism would imply strong competition between several
ordered phases, while the strong inhomogeneity in the system prevents formation of any long-range order.
As for the films, similar AFM fluctuations are expected, from which superconductivity emerges.
However, some factors could be unique to the films. First, with typical thickness
of $\sim$10~nm~\cite{Li_Nature_2019,Li_PRL_2020}, it should be easier to achieve
oxygen homogeneity upon annealing, and a more homogeneous environment helps acquiring the phase coherence for superconductivity.
Indeed, in superconducting Nd$_{0.8}$Sr$_{0.2}$NiO$_2$ films the resistivity remains metallic up to room temperature~\cite{Li_PRL_2020},
indicating a much higher level of electron homogeneity.
Furthermore, the SrTiO$_3$ substrate may provide additional doping or other interfacial effects
to suppress the static AFM order and induce superconductivity in thin films.
An example of the interface enhanced superconductivity has been realized in the single-layer
FeSe film grown on the SrTiO$_3$ substrates~\cite{Xue_CPL_2012,wang_scibull_2016}.

To summarize, we have performed $^1$H NMR measurements on the hydrogen-annealed powdered Nd$_{0.85}$Sr$_{0.15}$NiO$_2$ samples,
and observed enhanced AFM fluctuations below 100~K and emergence of a short-range AFM order
below 40~K. These properties are inherent to the strong correlations of Ni magnetic moments in Nd$_{0.85}$Sr$_{0.15}$NiO$_2$.
Although superconductivity and long-range AFM order are not observed,
likely due to the high inhomogeneity after annealing, the finding of strong AFM fluctuations pave the way
for the understanding of superconductivity in nickelates.

This work is supported by the National Natural Science Foundation of China with Grant Nos. 51872328, 11674392, 11774401 and A0402/11927809,
the Ministry of Science and Technology of China with Grant Nos. 2016YFA0300504 and 2016YFA0300401,
China Postdoctoral Science Foundation with Grant No. 2020M680797, the Fundamental Research Funds for the Central Universities,
and the Research Funds of Renmin University of China with Grant Nos. 18XNLG24, 20XNLG19 and 21XNLG18.

{

\end{document}